\documentstyle[aps,prl,epsfig]{revtex}

\begin{document}


\twocolumn[\hsize\textwidth\columnwidth\hsize\csname @twocolumnfalse\endcsname

\title{Dynamical interplay between local connectivity and global
competition\\ 
in a networked population}
\author{S. Gourley$^{1}$, S.C. Choe$^{1}$, P.M. Hui$^{2}$, and N.F. Johnson$^{1}$}
\address{$^1$ Clarendon Laboratory, Physics Department, Oxford University, Oxford
OX1 3NP, United Kingdom, \\
$^2$ Department of Physics, The Chinese University of Hong Kong, Shatin, New
Territories, Hong Kong}

\date{\today}
\maketitle

\begin{abstract} 
We show, both numerically and analytically, that the consequences of `wiring up'
a competitive population depend quite dramatically on the interplay between the
local connectivity and the global resources. With modest global resources,
adding small amounts of local connectivity yields an increasingly heterogeneous
population. With substantial global resources, high-performing yet reasonably
homogenous collective states emerge instead.  

\vskip0.2in
\end{abstract} ]

\vskip0.2in

Is `getting connected' a good or bad thing? How does increased
access to both global (i.e. public) information {\em and} local (i.e. private)
information, affect both the population as a whole and its individual
members? 
Thinking in terms of future technologies, what are the possible benefits and
dangers of introducing communication channels between collections of intelligent
devices, microsensors, semi-autonomous robots, nanocomputers, and even biological
micro-organisms such as bacteria \cite{economist,info}? These questions are
relevant to a wide range of computational, technological, biological and
socio-economic systems. However such questions are very difficult to
answer since these systems are simply `too complex'
\cite{stiglitz,arthur,challet1,anghel,johnson}. 

Here we
examine these questions using a minimal generic model which mimics a group of
selfish, rational individuals or components, referred to simply as `agents'. These
agents repeatedly compete for a global resource by making inductive decisions about
future group behavior.   The population and its agents may
be biological (e.g. a population of cellular organisms competing for nutrients),
computational (e.g. a grid of software modules
competing for processing time), mechanical (e.g. a constellation of sensors or
devices competing for communications bandwidth or operating power) or social (e.g.
a population of companies competing for business).  We uncover a rich  interplay
between the global competition for resources and the local inter-agent
connectivity. In a population with modest resources, low levels of
inter-connectivity increase the disparity between successful and unsuccessful
agents. By contrast in a higher-resource population with low inter-connectivity,
high-performing collective states can spontaneously arise in which nearly all
agents are reasonably successful. At high levels of inter-connectivity, the
overall population becomes fairer (i.e. smaller spread in success-rates) but less
efficient (i.e. smaller mean success-rate) irrespective of the global resource
level.

Our B-A-R (Binary Agent Resource) model is a minimal generic model of a
competitive networked population, which is based on Arthur's so-called El Farol
Problem \cite{arthur} in which a population of agents repeatedly
compete for a limited resource (i.e. seating in a crowded bar) \cite{details}.
Despite its everday human setting, the El Farol Problem 
embodies key generic characteristics of Complex Systems \cite{arthur,info}:
feedback and adaptation at the macroscopic and/or microscopic level, many
interacting parts, non-stationarity, evolution, coupling with the environment, and
observed dynamics which depend upon the particular realization of the system.
Although the El Farol Problem is more general than, say, the Minority
Game \cite{challet1,info,anghel} in that it allows for arbitrary levels of global
resource and is non-binary, the agents still only have access to {\em global}
information. Most biological, informational and socio-economic systems have at
least some degree of underlying connectivity among agents \cite{info} leading to
exchange of {\em local} information. It is the resulting interplay of network
connectivity and competition for global resources that we wish to
explore and understand. Therefore in contrast to previous works
\cite{arthur,challet1,anghel,info} our B-A-R model features network connectivity
{\em and} a general global resource level
$L$. Here $L$, which is not announced to the agents,  may represent the optimal
load capacity of a regional power grid, public utility, or electronic microcircuit;
the available space in a given urban region, public area or popular bar; the data
capacity of a given communications link in an informational or biological
system; the vehicular or passenger capacity on a given road or transport
system.

At each
timestep
$t$, each agent decides whether to access resource $L$  (i.e. action
$+1$ which might correspond to an
electronic component or device deciding to draw power, a computer sending a
data-packet down a given route, a surfer accessing a particular website, a commuter
taking a given road to work) or not to access this resource (i.e. action
$-1$). The two global outcomes at each timestep, `resource over-used' and
`resource not over-used', are denoted as `0' and `1'. If the number of agents
$n_{+1}[t]$ choosing action
$+1$ exceeds $L$ (i.e. resource over-used and hence global outcome `0') then the
$n_{-1}[t]$ abstaining agents win. By contrast if
$n_{+1}[t]\leq L$ (i.e. resource not over-used and hence global outcome `1') then
these $n_{+1}[t]$ agents win. Each agent decides its actions in light of (i) {\em
global information} which takes the form of the history of the $m$ most recent
global outcomes, and (ii) {\em local information} obtained via network
connections. Adaptation is introduced by randomly assigning $S$ strategies to each
agent. Each of the
$2^{2^m}$ possible strategies is a bitstring of length $2^m$ defining an action
($+1$ or $-1$) for each of the $2^m$ possible global outcome histories. For
example, $m=2$ has $2^2=4$ possible global outcome histories: $00$, $01$, $10$ and
$11$. Consequently there are $2^{2^{m=2}}=16$ possible strategies. Strategies
which predicted the winning (losing) action at a given timestep, are assigned
(deducted) one point. At each timestep, each agent compares the score of his own
best-scoring strategy (or strategies) with the highest-scoring strategy (or
strategies) among the agents to whom he is connected. The agent adopts the action
of whichever strategy is highest-scoring overall, using a coin-toss to break any
ties.  For simplicity we here assume a random network, 
where the connection
between any two agents exists with a probability
$p$. We emphasize that {\em any}
of the above model assumptions can easily be generalized: the numerical results
are reasonably robust.

Figure 1 shows the mean success-rate per agent, averaged over many numerical
realizations of the underlying connections and strategy distributions  (left axis:
solid curve) and the success-rate distribution within the population for a typical
run (right axis: histograms) as a function of the inter-connectivity
$p$ in a modest-resource population. We have chosen 
$L=(N-1)/2$, which is representative of modest resources since there are
more losers than winners.  We focus on the small
$m$ or `crowded' regime \cite{challet1,johnson,details} since there are relatively
few strategies compared to the number of agents and hence many agents will
simultaneously be competing to win with the same strategy.  The mean
success-rate decreases rapidly as $p$ increases  (left axis: solid curve) before
saturating  at
$p=p_{sat}\sim 0.1$. As
$m$ increases, $p_{sat}$ increases -- full results for the uncrowded, large $m$
regime will be discussed elsewhere.  The success-rate distribution for a typical
run (right axis: histogram) at $p=0$ exhibits a definite `class structure' in
terms of success. Dramatic changes then arise  as
$p$ increases. The spread in the success-rate   --
shown explicitly in the inset and by vertical arrows in Fig. 1 --
indicates a rapid increase in the population's heterogeneity in the range
$0
\leq p
\leq 0.02$. The success-rate distribution becomes almost continuous, washing out
the $p=0$ class structure. Above $p\sim 0.02$, there is a rapid drop in the
proportion of highly-successful agents. Further increasing $p$ leads to a decrease
of the spread in success-rates.  High levels of inter-connectivity therefore
provide increased fairness (i.e. small spread in success-rate) but decreased
efficiency (i.e. small mean success-rate).

Figure 2 compares our analytical results using the Crowd-Anticrowd theory
\cite{johnson,details} to the numerical results for the mean
success-rate per agent.  As will be explained in more detail elsewhere, the mean
success-rate per agent can be obtained from the standard deviation of the
number of agents making a given choice (e.g.
$+1$), by averaging this quantity over the attractors of the system (i.e.
averaging over the Eulerian Trail \cite{eulerian} or subsets of
it).  The mean success-rate can hence be expressed \cite{johnson,details} in terms
of the  mean number or {\em Crowd} of agents using the
$K$'th highest-scoring strategy 
(i.e. ${{n^{\rm net}_K}}$) and the mean number or {\em Anticrowd} of agents
using strategy $\overline K$ which is anticorrelated to $K$ (i.e.
${{n^{\rm net}_{\overline K}}}$ where ${\overline K}=2^{m+1}+1-K$).
Explicitly
\begin{equation} 
{ {n^{\rm net}_K}}={
{n_{K}+n_{\rightarrow K}-n_{K\rightarrow}}}
\end{equation}
where ${ {n_K}}$ is the number
of agents who would have used strategy $K$ in the absence of the
network because of the initial random strategy allocation:
\begin{equation} { {n_K}} = N\left( \left[ 1-\frac{(K-1)}{2^{m+1}}\right]
^{S}-\left[ 1-
\frac{K}{2^{m+1}}\right] ^{S}\right) \ .
\end{equation} 
In short,
${n_K}$ is the intrinsic Crowd-size due to crowding in the global strategy
space.
$n_{\rightarrow K}$ is a sum
over all agents who use strategy $K$ as a direct result of a network
connection:
\begin{equation} n_{\rightarrow K}=\bigg[\sum_{J>K}
{{n_J}}\bigg]\bigg[(1-p)^{\sum_{G<K} { {n_G}}}\bigg]\
\bigg[1-(1-p)^{{ {n_K}}}\bigg]\ \ .
\end{equation} 
Hence $n_{\rightarrow K}$ represents the {\em increase} in
Crowd-size due to the local connectivity. By contrast, $n_{K\rightarrow}$ is a
sum over all agents who would have used $K$ in the absence of a network, but who
will now use a better strategy as a result of their network connections:
\begin{equation} n_{K\rightarrow }={ {n_K}} \bigg[1-(1-p)^{\sum_{G<K}
{ {n_G}}}\bigg] \ \ .
\end{equation} 
Hence $n_{K\rightarrow}$ represents the {\em decrease} in
Crowd-size due to the local connectivity.  
The resulting analytic expressions for the mean success-rate are in
excellent agreement with numerical results (see Fig. 2) implying that the
population's dynamical evolution is indeed governed by crowds resulting
from the interplay between {\em local} connectivity and {\em
global} competition.

Figure 3 shows the mean success-rate per agent (thin solid lines)
as a function of the inter-connectivity $p$ in higher-resource
populations, for a range
of $L$ values and small
$m$. If the global resource level $L$ exceeds a critical amount
$L_{crit}=3N/4$, the mean success-rate per agent can exhibit a
{\em maximum} at small but finite inter-connectivity, for the following reason.
When
$L>L_{crit}$ and $p=0$, the population inhabits a `frozen' state in the sense that
the global outcome is persistently 1, i.e.
$\dots 111111$. With $S=2$ as in Fig. 3, each agent has probability $1/4$ of
being assigned two strategies which both define action $-1$ following a string of
$m$ `1's. Thus approximately $1/4$ of the population always lose and $3/4$ of the
population always win, leading to $L_{crit}=3N/4 \approx 75$. The global resource
is therefore under-used at $p=0$ by
$\Delta L\approx (L - 3N/4)$ at each timestep. Therefore increasing $p$ away from
$p=0$ will benefit some of the less successful agents by connecting them up to
successful agents, thereby giving them access to 
$\Delta L$, until  a $p$ value is reached where the connectivity is sufficient to
break the outcome series of 1's.  These run-averaged results can be better
understood by  considering run-specific results for $L=100$ as an illustrative
case (scattered circles). Specific runs appear to aggregate into groups
having similar temporal dynamics and success-rate distributions --  these groups
correspond to different dynamical states of the system. The increase in the mean
success-rate at low
$p$, corresponds to the system following an {\em Efficient} State E 
which has an outcome series of 1's as discussed above. The success-rate
distribution in State E (inset) is characterized by groups of
persistent winners and losers. As $p$ increases further, the
outcome series for high
$L$ tends to move through a set of  Intermediate States which combine a low
spread in success-rate  with a high mean  (e.g. Intermediate States A and B).
In Intermediate State A, the least-successful agents (which in State E had zero
success)  now have a success-rate which {\em exceeds} the average success-rate in
the high connectivity limit. In the high
$p$ limit, the outcome series tends toward the period-4 Eulerian Trail
\cite{eulerian} given by 
$\dots00110011\dots$. The corresponding number of agents taking action
$+1$ follows the pattern $\dots N,N,N/2,N/2,N,N,N/2,N/2 \dots$.  The
resulting {\em Fair} State F corresponds to all agents having access to the best
performing strategies, either by being assigned that strategy or by being
connected to another agent with that strategy. The resulting system is fair but
inefficient, having a small spread but also a small mean success-rate
$\approx 0.25$. 

Figure 4 shows the mean success-rate per agent (thin solid lines) as a function of
the inter-connectivity $p$ in a high-resource population ($L=95$) for various
history bit-string lengths
$m$.  The Intermediate  States, characterized by a reasonably high mean 
success-rate and a reasonably small spread (see Fig. 3 inset), have an
increasingly dominant effect on the system's  behavior as
$m$ increases. For a particular value of $m$, increasing $p$ moves the system
toward cycles of increasing period and hence the fractions of $1$'s and
$0$'s in the output series become more equal. The resulting State F, which
represents an increasingly noisy version of the Eulerian Trail as $m$ increases,
is fair but inefficient (see Fig. 3 inset).

We have derived analytic expressions (thick solid lines in Figs. 3 and 4) which
describe well the three dynamical regimes exhibited by the numerical results.
The  upper analytic branch at low
$p$, describing the efficient State E, is given by:
\begin{equation}
\frac{3}{4}+\frac{1}{4}\Bigg[1-(1-p)^{\frac{3N}{4}}\Bigg]\ \ .
\end{equation} 
The  middle analytic branch at intermediate $p$, describing the
Intermediate States, is given by:
\begin{equation}
\frac{3}{4}-\frac{9}{32}(1-p)^{\frac{7N}{16}}-\frac{1}{32}(1-p)^{\frac{15N}{16}}+
\frac{1}{8}(1-p)^{\frac{3N}{4}}\ \ .
\end{equation}
The  lower analytic branch at high $p$, describing the fair
State F, is given by:
\begin{equation}
\frac{1}{4}+\frac{9}{128}(1-p)^{\frac{7N}{16}}+\frac{1}{128}(1-p)^{\frac{15N}{16}}+
\frac{1}{16}(1-p)^{\frac{3N}{4}}\ \ .
\end{equation} 
The outcome series of 1's at low $p$ which yield
State E, can persist up to
$p_{crit}\sim 1 - (1 - 4\Delta L/N)^{4/3N}$. For $L=80,90,95$ and $100$,
this yields
$p_{crit}
\sim 0.002,0.011,0.019$ and $0.042$ respectively, which are also all in excellent
agreement with the numerical results. A 
full analysis of the dynamics of, and switching between, these regimes will be
given elsewhere.

In conclusion, we have reported a rich dynamical interplay between local
connectivity and global competition in a generic networked population. 
Apart from the intrinsic interest regarding functionality in complex system
networks, our results suggest that the internal
connectivity and global resources in such systems can be engineered in order to
enhance performance.

PMH acknowledges support from the Research Grants Council of the Hong Kong SAR
Government (grant CUHK4241/01P).

\newpage

\onecolumn

\noindent
\begin{figure}
\includegraphics{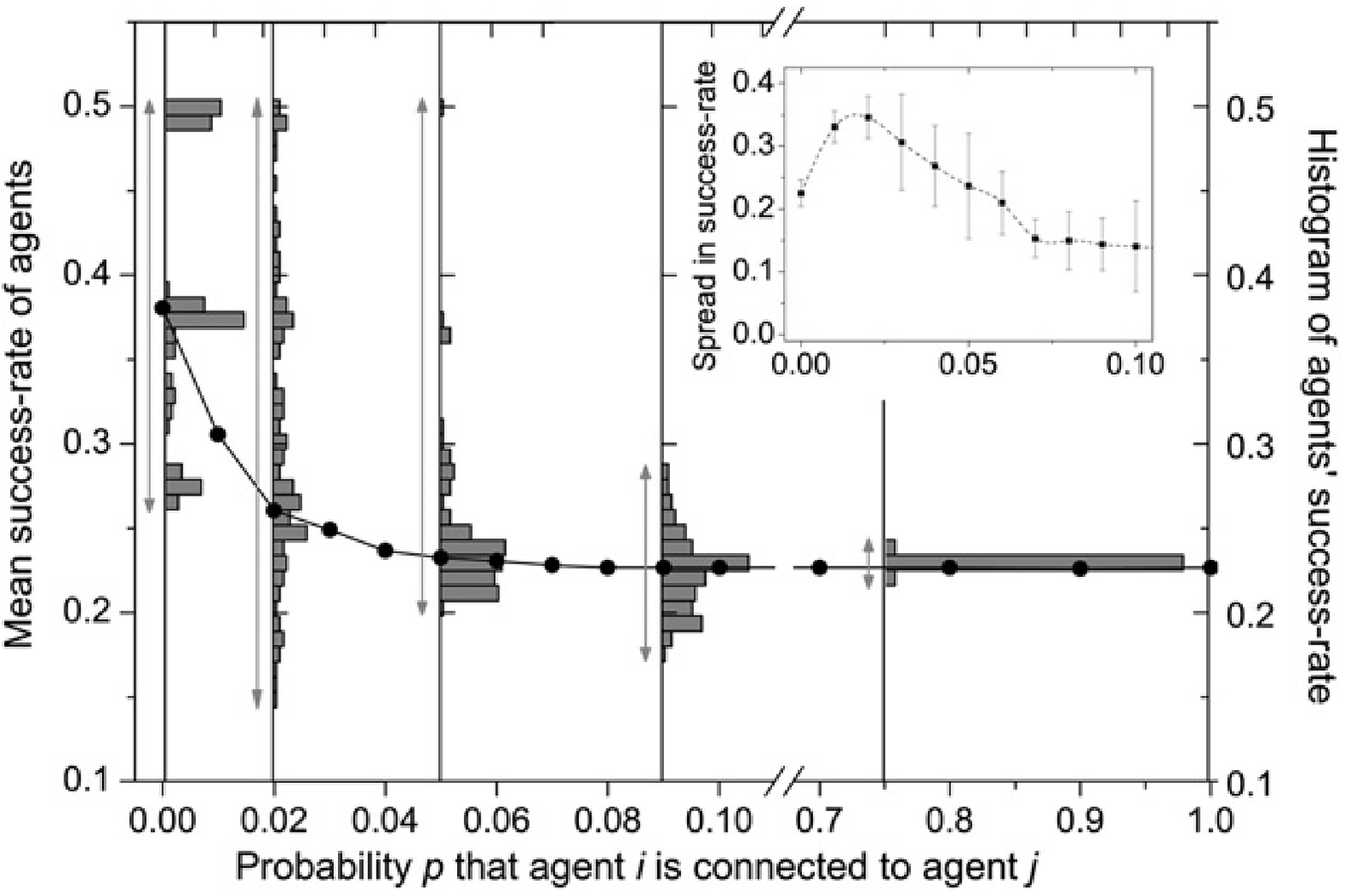}%
\vskip0.1in
\caption{Modest resource population: Solid line (left axis) shows
numerically-obtained mean success-rate per agent,  
as a function of the probability $p$ that agent $i$ is
connected to agent $j$. Right
axis: typical histograms of agents' success-rate in a typical run of
$10^{5}$ timesteps. $N=101$, $S=2$, $m=1$ and $L=50$. Inset: spread in
success-rate as a function of
$p$. Error bars obtained from 20 separate runs.}
\label{figure1}
\end{figure}

\newpage

\noindent
\begin{figure}
\includegraphics{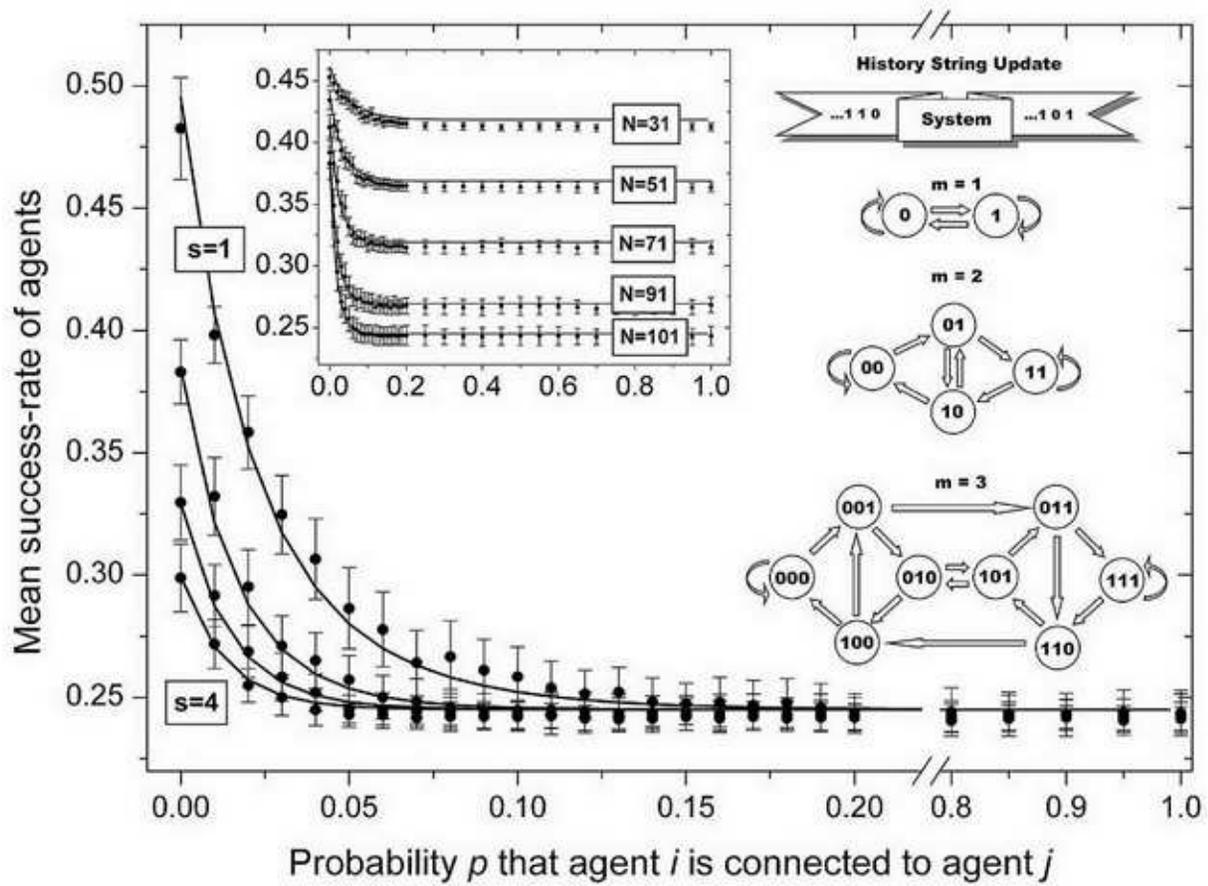}%
\vskip0.1in
\caption{Solid lines show analytic 
Crowd-Anticrowd theory for mean success-rate per
agent in a modest resource population, for $S=1,2,3,4$. Data-points are
numerical results. 
$m=1$, $N=101$, $L=50$. Inset: $S=2$, various $N$ values.
Right: Eulerian Trails [9] (i.e. high $p$ attractor) for $m=1,2,3$.}
\label{figure2}
\end{figure}

\newpage

\noindent
\begin{figure}
\includegraphics{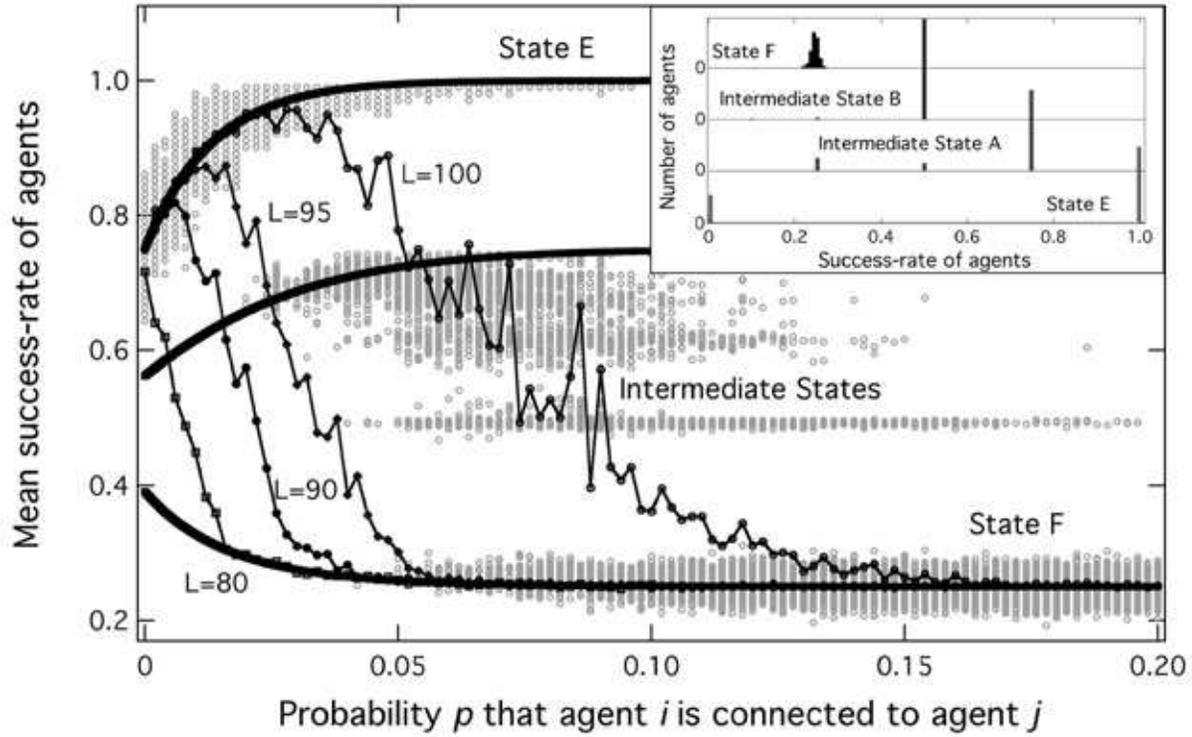}%
\vskip0.1in
\caption{Higher-resource population: Thin solid lines: numerically-obtained
mean success-rate per agent as a function of
$p$, at resource levels $L=80,90,95,100$. Results averaged  over 200 runs
of
$10^{5}$ timesteps,  with
$N=101$, $S=2$, $m=1$. Scattered circles show mean success-rate per agent for
separate runs at $L=100$ with $m=1$, as an illustration. Thick solid
lines: analytical curves (Eqs. (5)-(7)) describing the three dynamical regimes. 
The results agree so well over some portions, that the analytical curves obscure
the numerical results. Inset: histograms of typical success-rate distribution,
for 
$L=100$,
$m=1$. }  
\label{figure3}
\end{figure}

\newpage

\noindent
\begin{figure}
\includegraphics{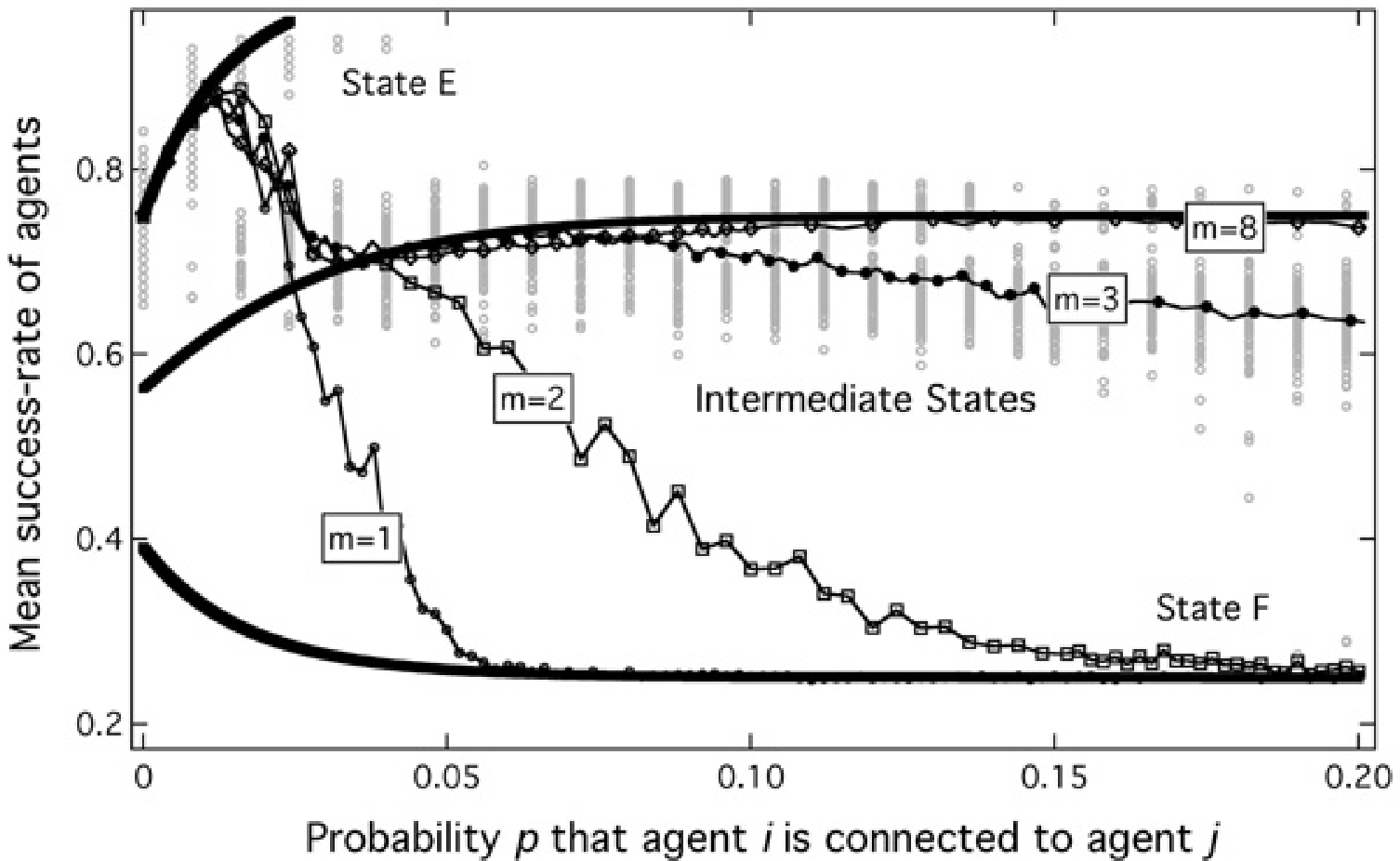}%
\vskip0.1in
\caption{High-resource population: Thin solid lines: numerically-obtained mean
success-rate per agent as a function of
$p$, at various $m$. Results averaged  over 200 runs of
$10^{5}$ timesteps,  with
$N=101$, $S=2$, $L=95$. Scattered circles show mean success-rate per agent for
separate runs at $m=3$. Thick solid
lines: analytical curves (Eqs. (5)-(7)) describing the three dynamical regimes. 
The results agree so well over some portions, that the analytical curves obscure
the numerical results.}
\end{figure}


\begin{thebibliography}{99}
\bibitem{eulerian} The Eulerian Trail corresponds to a particular cycle around the
de Bruijn graph in which every possible transition between global information
states is visited once. The de Bruijn graph  represents the
$2^m$ possible global information states.
\end{thebibliography}
\end{document}